# Optomechanical Measurement of the Abraham Force in an Adiabatic Liquid Core Optical Fiber Waveguide

H Choi[1], M Park[1], D S Elliott[2], and K Oh[1]

[1]*Department of Physics, Yonsei University, 50 Yonsei-ro Seodaemun-gu, Seoul, 120-749, Korea*
[2]*School of Electrical and Computer Engineering and Department of Physics and Astronomy, Purdue University, West Lafayette, Indiana, 47907-2035, USA*

E-mail: koh@yonsei.ac.kr

**Abstract.** We report quantitative experimental measurements of the Abraham force associated with a propagating optical wave. We isolate this force using a guided light wave undergoing an adiabatic mode transformation (AMT) along a liquid-filled hollow optical fiber (HOF). Utilizing this light intensity distribution within the liquid, we were able to generate a time-averaged non-vanishing Abraham force density, while simultaneously suppressing the Abraham–Minkowski force density. The incident laser field induced a linear axial displacement of the air–liquid interface inside the HOF, which provided a direct experimental measure of the Abraham force density. We find good agreement between the experimental results and theoretical determinations of the Abraham force density.

## 1. Introduction

The momentum carried by an electromagnetic field in a dielectric medium has been the subject of simmering debate since the early 1900's. Minkowski [1] proposed that this momentum is of the form $\boldsymbol{D} \times \boldsymbol{B}$, while Abraham [2] advocated the form $c^{-2} \boldsymbol{E} \times \boldsymbol{H}$. Over the decades, a number of works [3-10] have provided support of one model or the other. For reviews of these works, see Brevik, 1979; Baxter and Loudon, 2010; and Milonni and Boyd, 2010 [11-13]. In vacuum, of course, these expressions are equivalent. In dielectric media, however, these expressions differ, and their difference leads to a small, but observable, force density, known as the Abraham force density, which can be exerted by an electromagnetic field on the medium. Recently Barnett [3] proposed that both forms of the momentum density are in fact correct, with Minkowski giving the canonical momentum, and Abraham the kinetic momentum. While the total momentum of the system (light plus medium) must, of course, be unique, assignment of the momentum to the light field or the medium through which it propagates is not, and this distinction separates the two descriptions of the momentum. The growing acceptance of this interpretation of the momentum densities [12,13] and resolution of the Abraham-Minkowski debate does not diminish the impact of a direct observation of the Abraham term exerted by propagating electromagnetic waves, which has been a long-sought, but previously unsatisfied, goal. In this work, we report observation and measurement of the Abraham force density using optical waves undergoing an adiabatic mode transformation (AMT) in a liquid-filled, hollow-core optical fiber.

In the Abraham formalism, the electromagnetic force density in a nonmagnetic dielectric medium with the refractive index of $n$ is given as [11]:

$$\boldsymbol{f} = \boldsymbol{f}^{AM} + \boldsymbol{f}^{A} = -\frac{\varepsilon_0}{2} \boldsymbol{E}^2 \nabla n^2 + \frac{n^2-1}{c^2} \frac{\partial}{\partial t}(\boldsymbol{E} \times \boldsymbol{H}). \tag{1}$$

The first and second terms in this expression are $\boldsymbol{f}^{AM}$, the Abraham–Minkowski force density, and $\boldsymbol{f}^A$, the Abraham force density. From this form of $\boldsymbol{f}^A$, we see that the Abraham force density vanishes for a time-harmonic, one-dimensional wave. Thus the Abraham force density has been observed unambiguously only for fields that are essentially static [14-16]. A clever scheme [17] for the observation of the Abraham force for such harmonic waves has been proposed, but to date none have been reported. An additional impediment to a clean measurement of $\boldsymbol{f}^A$ is the Abraham-Minkowski force density $\boldsymbol{f}^{AM}$, which in most geometries is much larger than $\boldsymbol{f}^A$ [12]. As we discuss below, $\boldsymbol{f}^{AM}$ is suppressed in the present measurement by the exchange of electromagnetic momentum without significant refractive index inhomogeneity.

## 2. The non-vanishing Abraham force density and the Abraham-Minkowski force density

The propagating electromagnetic modes in a hollow optical fiber (HOF) have been studied in great detail [18, 19]. When one end of such a fiber is filled with a liquid of refractive index $n$, the optical mode transforms from the ring region of the air-filled hollow fiber to the liquid-filled central region. This mode transformation is adiabatic and well characterized. As we will show in this paper, the Abraham force density in this system, while small, is much greater than the Abraham-Minkowski force density, and is directly observable through measurements of the axial displacement of the liquid as it is drawn into the hollow-core region. We have performed measurements of this liquid displacement at varying powers of the optical mode, using three different liquids of varying refractive index. We see excellent agreement between measurements and calculations of the magnitude of the Abraham force density.
.

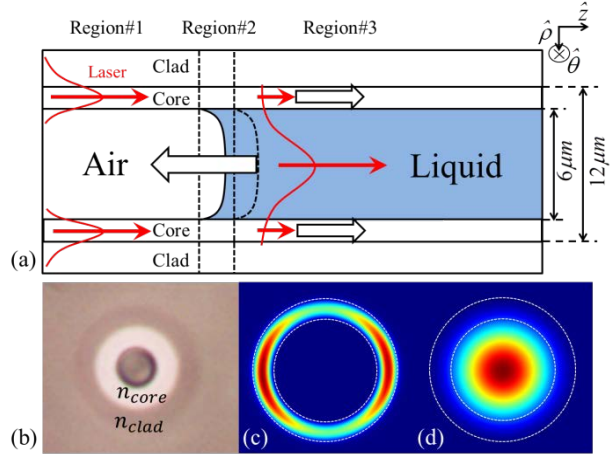

**Figure 1.** (a) The AMT of the optical wave occurs between regions #1, #2 and #3. The momentum density carried by the light (bold arrows), and the force density on the media (void arrows) are shown. The material force density will continuously displace the liquid when a cw laser beam is launched. (b) The cross section of the HOF where $n_{core}=1.4452$, and $n_{clad}=1.4402$ [20, 21]. The expected fundamental mode field distributions at (c) region #1 and (d) region #3, calculated by FEM. The white dotted lines are the boundaries shown in (b).

We show the experimental geometry schematically in figure 1(a). The HOF consists of a central air hole, germanium-doped silica ring core, and silica cladding [18-21]. The hole of the HOF was partly filled with a dielectric liquid whose refractive index is higher than that of the ring core to form a step index waveguide. The adhesive force between the liquid and the silica inner surface of the HOF resulted in a concave interface between the air and the liquid, which efficiently and adiabatically transforms the optical mode from the ring to the liquid core along the HOF. The laser mode displaced the position of the air-liquid interface toward the laser launching side, which provided a means for the experimental isolation and measurement of the Abraham force.

The key features of the measurement of the Abraham force are (1) the optical mode supported by the HOF is transformed along the three segments of the fiber [the ring core in the air-filled HOF (region #1); the air-liquid interface (region #2); and the liquid core in the liquid-filled HOF (region #3)]; (2) the momentum carried by the optical wave is primarily in the ring in region #1, but is adiabatically transformed to the liquid-filled core in region #3, resulting in a non-vanishing, time-averaged Abraham force felt by the liquid; (3) the amplitude of the fundamental mode at the air-liquid interface and the ring core and liquid interface are negligible, reducing the Abraham-Minkowski force density to levels much less than the Abraham force density; and (4) the displacement of the air-liquid interface along the axial direction of the HOF is linear in the incident laser power, allowing a direct measurement of the Abraham force. Our physical picture for the liquid displacement is rooted in its recoil as the light, which enters the liquid region from the side, is deflected into line with the fiber axis. Although the Abraham force density has a large time-harmonic component (whose average value vanishes), there remains a measureable, non-vanishing quantity related to the variation of the Poynting vector during the mode transformation to the liquid-filled core. This is most directly seen through the following analysis.

In order to determine the dependence of the magnitude of the Abraham force density on the laser power, we analyze the time derivative term in equation (1) as follows [13]:

$$\frac{\partial}{\partial t}(\mathbf{E} \times \mathbf{H}) = \frac{\partial \mathbf{E}}{\partial t} \times \mathbf{H} + \mathbf{E} \times \frac{\partial \mathbf{H}}{\partial t} = \frac{1}{\varepsilon}(\nabla \times \mathbf{H} - \mathbf{J}) \times \mathbf{H} + \mathbf{E} \times \left(-\frac{1}{\mu}\nabla \times \mathbf{E}\right), \quad (2)$$

where we have used Ampere's Law and Faraday's Law to eliminate the time derivatives of $\mathbf{H}$ and $\mathbf{E}$. Using common vector differential identities [22], this becomes

$$\frac{\partial}{\partial t}(\mathbf{E} \times \mathbf{H}) = \frac{1}{\mu_0}\left[(\mathbf{E} \cdot \nabla)\mathbf{E} - \frac{1}{2}\nabla \mathbf{E}^2\right] + \frac{1}{\varepsilon}\left[(\mathbf{H} \cdot \nabla)\mathbf{H} - \frac{1}{2}\nabla \mathbf{H}^2\right], \quad (3)$$

for a charge-free, dielectric medium. Equation (3) results in an Abraham force density along the z-direction of an isotropic medium of the form

$$f_z^A = \frac{n^2-1}{c^2}\sum_{i=1}^{3}\left\{\frac{1}{\mu_0}\frac{\partial}{\partial x_i}\left[E_z E_i - \frac{1}{2}\delta_{zi}\mathbf{E}^2\right] + \frac{1}{\varepsilon}\frac{\partial}{\partial x_i}\left[H_z H_i - \frac{1}{2}\delta_{zi}\mathbf{H}^2\right]\right\}, \quad (4)$$

where the indices $i = 1, 2$, or $3$ represent the $x, y,$ or $z$ components of the various vector quantities. The total Abraham force acting on the liquid in figure 1(a) can be obtained by volume integration of equation (4) within the liquid volume between the regions #1 and #3. The $E_z$ field contribution in equation (4) is negligible for weakly-guided waves [23]. Then, by the divergence theorem, the volume-integrated Abraham force becomes:

$$\int_V f_z^A dV \approx -\frac{1}{2}\left[\int_{A_3}\frac{n^2-1}{n^2}[\mathbf{E}\cdot\mathbf{D}+\mathbf{B}\cdot\mathbf{H}]dA - \int_{A_1}\frac{n^2-1}{n^2}[\mathbf{E}\cdot\mathbf{D}+\mathbf{B}\cdot\mathbf{H}]dA\right], \quad (5)$$

where $A_1$ and $A_3$ are the cross-sectional areas of the inner hole of the HOF at regions #1 and #3, respectively. Since air (for which $n^2-1 \ll 1$) fills the hole in region #1, the surface integral at $A_1$ vanishes. The surface integral in equation (5) over $A_3$, however, remains, leading to a measurable, non-vanishing time-averaged contribution to the Abraham force density. Also, since the force in equation (5) is proportional to the integral of the energy density of the wave, the total Abraham force is expected to be proportional to the input power of the laser beam.

For unambiguous observation of the Abraham force density $f^A$, it is necessary to suppress the Abraham-Minkowski force density $f^{AM}$, which in most geometries is the dominant contribution. We demonstrate here that $f^{AM}$ is negligible in the hollow optical fiber geometry. The total Abraham-Minkowski force exerted on the liquid is

$$\bar{f}_z^{AM} = \int_V f_z^{AM} dV = -\frac{\varepsilon_0}{2} \int_V |\boldsymbol{E}|^2 \frac{\partial n^2}{\partial z} dV. \tag{6}$$

Here the volume $V$ is the cylindrical volume shown as region #2 in figure 1, which includes the air-liquid interface. We used a finite element method (FEM, COMSOL Multiphysics) to analyze the fiber modes numerically, obtaining the electromagnetic field distribution at the concave interface. In order to determine the electric field amplitude at the interface, we modeled the concave air-liquid interface at region #2 as a stack of ring-shaped liquid layers within the ring core in 0.3 μm increments. Using this analysis, we calculated the total Abraham-Minkowski force ($\bar{f}_z^{AM}$), which we compared with the total Abraham force ($\bar{f}_z^A$). We present these results in Table 1, in which each force is normalized to the optical power of the fiber mode. We chose the refractive indices of liquids similar to the index of ring core to suppress multimode propagation. We assumed that the propagation mode is the fundamental mode, which is approximately valid for the liquid refractive indices of n = 1.4512 and 1.4604, but less so for n = 1.4712, as we discuss later. The uncertainties of the calculated time-averaged total Abraham force listed in this table are derived only from the refractive index uncertainty ($\pm 0.0002$) of the liquids [24].

**Table 1.** Results of the numerical calculations of total $\bar{f}_z^{AM}$ and the total $\bar{f}_z^A$.

| Refractive index | $\bar{f}_z^{AM}/P_{laser}$ (fN/mW) | $\bar{f}_z^A/P_{laser}$ (pN/mW) | $\bar{f}_z^{AM}/\bar{f}_z^A$ |
|---|---|---|---|
| 1.4512 | 35.0 | 1.26 ± 0.03 | 0.027 |
| 1.4604 | 86.6 | 2.06 ± 0.01 | 0.042 |
| 1.4712 | 195 | 2.39 ± 0.01 | 0.081 |

As shown in this table, $\bar{f}_z^{AM}$ is in each case much less than $\bar{f}_z^A$. We understand this result through examination of equation (6). The Abraham-Minkowski force density exists where the refractive index inhomogeneity differs from zero ($\nabla n^2 \neq 0$), which is the case at the air-liquid interface. In this region, however, the intensity of the optical mode is strong in the ring, rather than in the central hole. Based on this analysis, therefore, we are justified in ignoring the Abraham-Minkowski force $\bar{f}_z^{AM}$ on the liquid, and interpret the displacement of the liquid in the hollow fiber as a determination of the total Abraham force $\bar{f}_z^A$.

## 3. Experimental setup

We show a schematic of the experimental setup in figure 2(a). The output of the tunable laser source (TLS, Agilent, N7714A) is at a wavelength of $\lambda = 1550\ nm$. At this wavelength, the hollow optical fiber has its lowest attenuation. The light from the laser source formed a fundamental mode, and was delivered using single-mode fiber (SMF, Corning SMF-28). We amplified this laser output using an erbium-doped fiber amplifier (EDFA, FiberLabs, AMP-FL8013-CB, pumped by a 980 nm laser diode) to a power in the range 20-100 mW (in ~10 mW increments), which we controlled by changing the drive current level. Due to splicing losses (~*25%*), however, the laser power delivered to the hollow optical fiber ranged from 15-75 mW. In order to suppress the thermal effects at the splicing point between SMF and HOF, we immersed this fiber junction in an ethanol reservoir. We measured the total output power at the end of the HOF. Light absorption by the liquid was less than the sensitivity of our power meter (<1 dB/cm). We measured the air-liquid interface position using an optical microscope with a x40 objective lens (OBJ, Leica, FL-Plan) and a CCD camera. We filled three HOFs prepared to a uniform length ( $40 \pm 0.3$ cm ) with different dielectric liquids (Cargille Series A), which had refractive indices of $n_{liq}$=*1.4512*, *1.4604*, and *1.4712* at $\lambda$=*1550 nm* (with the index uncertainty of ±*0.0002* [24]).

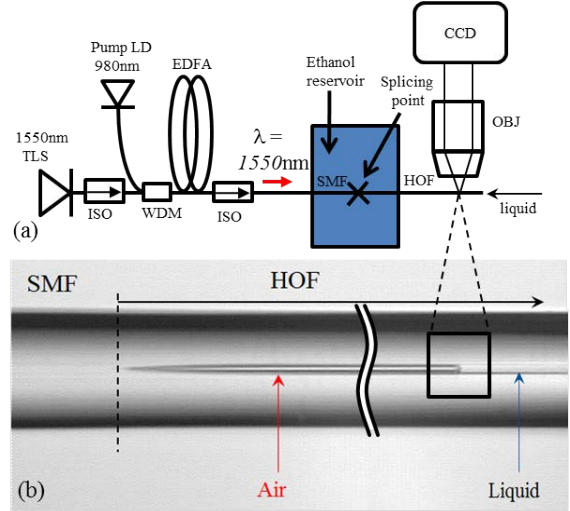

**Figure 2.** (a) Schematic of the experimental set up. (b) An actual photograph of the liquid filled HOF.

We show in figure 2(b) a microscope image of the liquid-filled hollow fiber. The fiber was deployed horizontally to minimize gravitational effects. At the left side, the fiber was spliced to the single-mode fiber, while at the right side, the liquid filled the hollow central region. In the absence of the laser input, the equilibrium position of the air–liquid interface in the HOF was determined by the internal air pressure and the capillary force between the liquid and HOF. Before the experiment, we tested each of the HOFs by launching a cw laser beam with an incident power of 20 mW for 1 second. By this means, we rejected fibers that displayed excessively over-damped or under-damped behavior, related to surface adhesion between the liquid and the HOF, which would introduce errors into the measurements of the displacement. For measurements of the Abraham force, the interface shifted in the direction opposite to the laser propagation direction, and reached its equilibrium position within ~0.5 sec after application of the incident power. We show photographs of the air-liquid interface for $n_{liq}$ = *1.4604* in figure 3, clearly showing the shifts with various laser powers. We determined the positions of air-liquid interface using the Canny edge-detection algorithm [25]. The interface displacement $\Delta l_{eq}$ increased monotonically with the laser power. Due to the effects of surface adhesion, the reference position of the air-liquid interface when the laser power was turned off was difficult to determine to the same precision as measurements of the interface position with the laser turned on. (The final equilibrium position of the air-liquid interface depended slightly on whether the interface was shifting towards or away from the laser source. Thus, the reference position, which we measured by turning the laser off, causing the interface to the right in figure 2, had some built-in bias relative to the measurements of the displacement when we turned the laser power on, causing the interface to shift to the left.)

We also investigated the thermal-expansion-induced displacement of the interface by exposing the fiber to the cw laser output for longer periods. The thermal expansion of the internal air within the HOF can cause a shift of the reference position of the interface. For the laser power of *50 mW*, the system remained stable for one hour without detectable changes in $\Delta l_{eq}$, confirming that the shifts were not thermally induced in our experiments. For a higher laser power of *100 mW*, thermally-induced internal air expansion took a few minutes to reach the equilibrium position.

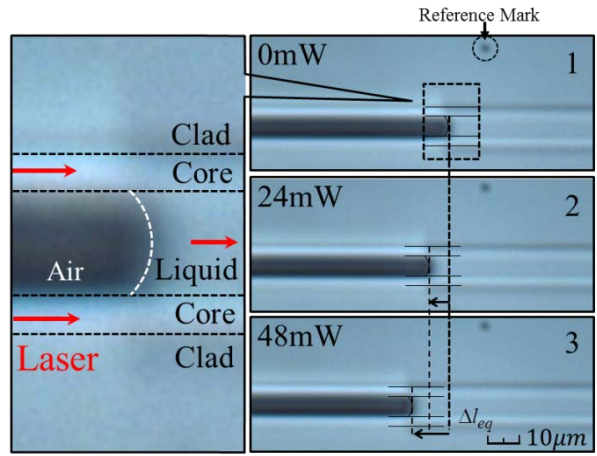

**Figure 3.** Images collected with the objective lens and CCD camera in figure 2 and processed by the Canny edge-detection algorithm [25]. The positions of the air–liquid interface were defined by the meniscus. For these images, $n_{liq}$ = 1.4604. The displacement was in the direction opposite the laser beam propagation direction, and $\Delta l_{eq}$ increased linearly with the laser power [26].

4. **Data Analysis**

By measuring $\Delta l_{eq}$ at each laser power, we determine the total Abraham force along the z-direction, $\bar{f}_z^A$, using the pressure equilibrium condition [27]:

$$\frac{\bar{f}_z^A}{A_3} \approx \left(P_{ATM} + P_{cap}\right)\frac{\Delta l_{eq}}{l_{HOF}}, \quad (7)$$

where $A_3$ is the area of the HOF's central hole, $l_{HOF}$ is the length of the HOF, and $P_{ATM}$ and $P_{cap}$ are the atmospheric pressure and the capillary-force-induced pressure of the liquid on the inner surface of the HOF, respectively. Here we assumed a constant and uniform room temperature in the system and $\Delta l_{eq} \ll l_{HOF}$ as indicated by figure 2 and 3. We determined the internal air pressure ($P_{ATM} + P_{cap}$) by measuring the length of the air region in the HOF; its average value was $107.98 \pm 0.03 \, kPa$, which is slightly greater than the atmospheric pressure by the capillary force. We present the experimentally-determined total Abraham force, $\bar{f}_z^A$, as the data points in figure 4. Each data point represents a single measurement. The magnitude of the force is of the order of pico-newtons, increasing linearly with incident laser power, consistent with equation (5).

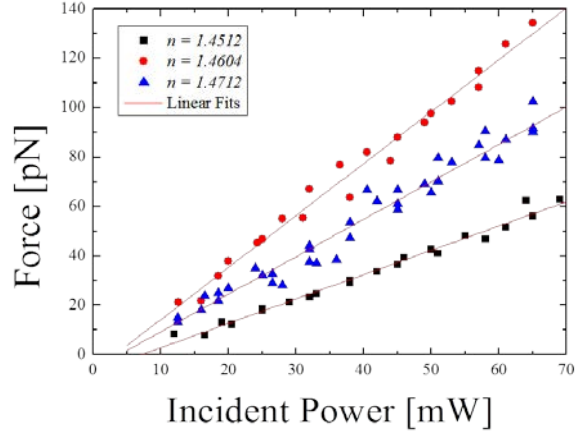

**Figure 4**. Plot of the Abraham force vs the power $P_{laser}$ of the laser beam, for three liquids with various refractive indices. We determine the force $\bar{f}_z^A$ from the displacement $\Delta l_{eq}$ of the liquid, using equation (7).

We fit each of these data sets to a straight line to determine $\bar{f}_z^A / P_{laser}$ for each refractive index liquid. We found that the slope $\bar{f}_z^A / P_{laser}$ changed significantly with $n_{liq}$, and summarize the results in Table 2. The slope does not scale with the refractive index, which is further evidence that the Abraham-Minkowski force is only a minor contributor to the total force in this work.

**Table 2.** The ratio of the total Abraham force applied to each of the liquids to the laser power, as determined experimentally (Exp) and numerically (FEM).

| $n_{liq}$ | $\bar{f}_z^A / P_{laser}$ (Exp.) *(pN/mW)* | $\bar{f}_z^A / P_{laser}$ (FEM) *(pN/mW)* |
|---|---|---|
| 1.4512 | $0.96 \pm 0.03$ | $1.15 \pm 0.03$ |
| 1.4604 | $2.10 \pm 0.06$ | $2.01 \pm 0.01$ |
| 1.4712 | $1.51 \pm 0.05$ | $1.74 \pm 0.01$ |

The fitted intercepts for each data set shown in figure 4 are $-7.0 \pm 1.2 \, pN$, $-6.7 \pm 2.6 \, pN$, and $-5.9 \pm 2.1 \, pN$, respectively. These intercepts are consistent with one another, and are consistent with the expected precision of our measurement of the reference position of the air-liquid interface with the laser power turned off, discussed above.

5. **Analysis**

As presented in equation (5), a proper analysis of the Abraham force requires a detailed characterization of the optical mode pattern within the liquid-filled region of the hollow optical fiber. The ring-shaped mode of region #1, shown in figure 1(c), transforms into the filled-in distribution in region #3, shown in figure 1(d). In our previous discussion in Section 2, we obtained the Abraham force density, but limited that analysis to include only the fundamental mode. A variety of modes are excited, however, and we require the relative amplitude of each mode to estimate precisely the total force. The general electric field intensity in terms of the multiple fiber modes is given by

$$I(x,y,z) = \sum_{m,n=1}^{\infty} E_m(x,y) E_n(x,y) \exp\left[-i(\beta_n - \beta_m)z\right], \quad (8)$$

where $E_m(x,y)$ are the mode amplitudes, and the propagation constant of the $m$th mode $\beta_m = 2\pi n_{eff} \lambda^{-1}$ depends on an effective refractive index ($n_{eff}$) of the mode. In order to obtain the multimode interference within the liquid, we used a numerical software package (Rsoft, Photonics CAD ver 5.1.8), assuming that the field amplitudes vary slowly compared to $exp(-i\beta z)$ [28, 29]. We show a color map of the mode pattern for the n = 1.4512 liquid in figure 5(a).

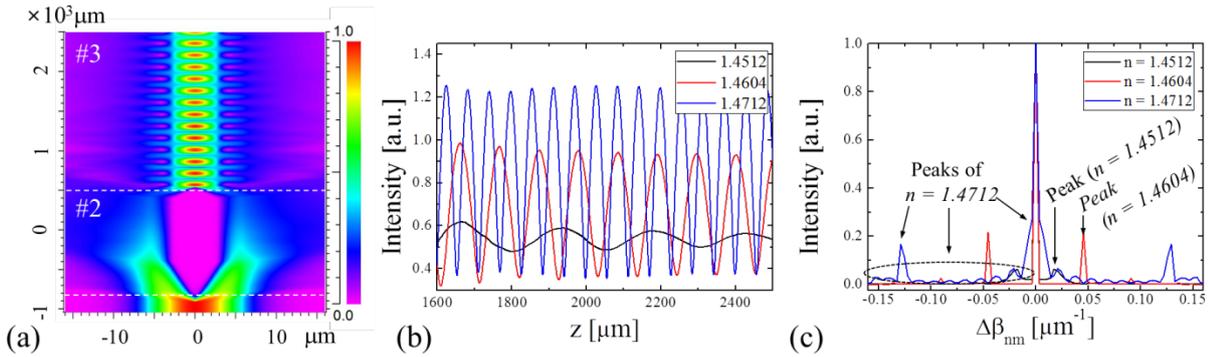

**Figure 5**. (a) The mode pattern for the case of the *n = 1.4512* liquid. The light enters from the bottom in the single-mode fiber (SMF), transforming to the filled-in distribution in the liquid-filled hollow optical fiber. The number #2 and #3 represent the region #2 and #3 in figure 1(a) respectively. (b) The intensity pattern at the center of HOF for *n = 1.4512* (black), *n = 1.4604* (red), and *n = 1.4712* (blue) vs *z*, the axial distance along the fiber. (c) The Fourier transformed results of (b). Note the highly multimode characteristics for *n = 1.4712*.

The electric field pattern along the fiber axis obtained by this method is shown in figure 5(b). This plot illustrates the modulation in the net field amplitude, due to the interference between the different modes, which is of greatest magnitude for the case of the n = 1.4702 liquid. We Fourier transform this calculated field pattern, shown in figure 5(c), in order to extract the amplitudes $E_m$ and propagation constants $\beta_m$ of the individual modes. The Fourier transformation of equation (8) is given in momentum space as

$$\tilde{I}(k) = \frac{1}{2\pi} \sum_{m,n=1}^{\infty} E_m E_n \delta(k - \Delta\beta_{nm}), \quad (9)$$

where $\Delta\beta_{nm}$ is the difference between mode propagation constants, $\Delta\beta_{nm} = \beta_n - \beta_m$. The peak at $k = 0$ shows the sum of modal intensities, and the peaks at $k = \Delta\beta_{nm}$ show the amplitudes of $E_m E_n$ in equation (9). From the result of figure 5(c), we estimated $E_m E_n$ from the amplitudes of the peaks. The frequency positions of $\Delta\beta_{nm}$ were estimated by obtaining propagation constants of highly excited modes by FEM. The differences showed good agreement with the Fourier transform results in figure 5(c). For example, in the case of *n = 1.4512*, we have a peak at $\Delta\beta_{nm} = 0.01751$ by the Fourier transform results, while the difference between the propagation constants of the fundamental and first excited modes determined by FEM is $\Delta\beta_{nm} = 0.01832$. Here we determined that 92.3% of the optical power propagates in the fundamental, and 7.7% in the 1$^{st}$ excited mode. Also, in the case of *n = 1.4604*, we find $\Delta\beta_{nm} = 0.04546$ and $\Delta\beta_{nm} = 0.04084$ by Fourier transformation results and FEM, respectively. The fundamental and 1$^{st}$ excited mode carry 90.8% and 9.1% of the optical power, respectively. Thus, for n = 1.4512 and n = 1.4604, most of the power is in the fundamental mode, and only one excited mode has any significant excitation. In contrast, we find highly multimode behavior when *n = 1.4712*; with significant excitation of 8 modes. Strong peaks at $\Delta\beta_{nm} = 0.125$ and 0.005, and several weaker peaks can be seen in figure 5(c) for this case. We determined that 63.7% of the laser power propagates as the 2$^{nd}$ excited mode and 23.5% as the fundamental mode, with the remaining power distributed over several modes.

We calculated the total Abraham force for each of the three different refractive index liquids using equation (5), considering the excited mode contributions at region #3. We present these numerical results in Table 2 for comparison with the experimental results. The uncertainties in the measured results presented in this table are derived from the scatter in the data points in figure 4, while the uncertainties in the calculated results reflect only the mode amplitudes, as just discussed.

The differences between the experimental and numerical results in this table are less than 20% in each case, of order 0.1 pN/mW. While these differences are larger than the stated uncertainties, those uncertainties do not reflect differences between the fibers that can originate from the fabrication process. For example, there could exist a localized surface adhesion force difference, resulting from a variation in the temperature during the fabrication process [30]. During the experiment, we used various HOF samples, and it is not unreasonable that variations among the fibers could be responsible for the variations in the measured results of this magnitude.

## 6. Conclusion

In summary, we have experimentally demonstrated the isolation and measurement of the Abraham force of an optical wave on a dielectric medium by utilizing the AMT along a liquid-filled hollow optical fiber. Due to its waveguide properties along the liquid-filled HOF, the Abraham force is much larger than the Abraham-Minkowski force, and the linear displacement of the air-liquid interface along the axial direction as a function of the incident laser power allows its direct determination. We demonstrated that the Abraham force is proportional to the laser power. We also analyzed the Abraham force numerically, using FEM to determine the fundamental optical mode distributions, and showed good agreement with experimental results. Our observation does not prove the uniqueness of the Abraham's expression as a general expression [3,9]; rather it demonstrates that the Abraham force density for optical waves exists, and that it can be emphasized and measured under specific conditions.

**Acknowledgments**

We are grateful to J. Choi for helpful discussions. This work was partly supported by the ICT R&D program of MSIP/IITP [2014-044-014-002, Development of core technologies for quantum cryptography networking].

[1] Minkowski H 1908 Die Grundgleichungen fur die elektromagnetischen Vorgange in bewegten Korpern, Nachr. Konigl. Ges. Wiss. Goettingen Math. Phys. K1 53 [reprinted as 1910 Math. Ann. **68**, 472]
[2] Abraham M 1909 Zur Elektrodynamik bewegter Korper, Rend. Circ. Mat. Palermo **28** 1 ;1910 **30** 33
[3] Barnett Stephen M 2010 Resolution of the Abraham-Minkowski dilemma, Phys. Rev. Lett. **104**, 070401
[4] Ashkin A and Dziedzic J M 1973 Radiation pressure on a free liquid surface, Phys. Rev. Lett. **30**, 139
[5] Jones R V 1951 Radiation Pressure in a Refracting Medium, Nature (London) **167**, 439
[6] Jones R V and Leslie B 1978 The measurement of optical radiation pressure in dispersive media, Proc. R. Soc. London, Ser. A **360**, 347
[7] Jones R V and Richards J C S 1954 The pressure of radiation in a refracting medium, Proc. Roy. Soc. A **221**, 480
[8] Gordon James P 1973 Radiation forces and momenta in dielectric media, Phys. Rev. A **8**, 14
[9] Hinds E A and Barnett Stephen M 2009 Momentum exchange between light and a single atom: Abraham or Minkowski?, Phys. Rev. Lett., **102**, 050403
[10] She W, Yu J, and Feng R 2008 Observation of a push force on the end face of a nanometer silica filament exerted by outgoing light, Phys. Rev. Lett. **101**, 243601
[11] Brevik I 1979 Experiments in phenomenological electrodynamics and the electromagnetic energy momentum tensor, Phys. Rep. **52**, 133
[12] Baxter C and Loudon Rodney 2010 Radiation pressure and the photon momentum in dielectrics, J. of Mod. Opt. **57**, 830-842
[13] Milonni P W and Boyd R W 2010 Momentum of light in a dielectric medium, Adv. in Opt. and Phot. **2**, 519-553
[14] Walker G B, Lahoz D G and Walker G 1975 Measurement of the Abraham force in a barium titanate specimen, Can. J. Phys. **53**, 2577
[15] Walker G B and Walker G 1976 Mechanical forces of electromagnetic origin, Nature **263**, 401
[16] James R P 1968 Force on Permeable Matter in Time-varying Fields, Ph.D. Thesis, Dept. of Electrical Engineering, Stanford University
[17] Brevik I and Ellingsen S Å 2010 Possibility of measuring the Abraham force using whispering gallery modes, Phys. Rev. A **81**, 063830
[18] Choi S, Oh K, Shin W, Park C S, Paek U C, Park K J, Chung Y C, Kim Y, and Lee Y G 2002 Novel mode converter based on hollow optical fiber for gigabit LAN communication, IEEE Photonics Tech. Lett. **14**, 248
[19] Lee S, Park J, Jeong Y, Jung H, and Oh K 2009 Guided wave analysis of hollow optical fiber for mode-coupling device applications, J. Light. Tech. **27**, 4919
[20] Fleming J W 1984 Dispersion in $GeO_2$-$SiO_2$ glasses, Appl. Opt. **23**, 4486
[21] Oh K, Choi S, Jung Y, and Lee J W 2005 Novel hollow optical fibers and their applications in photonic devices for optical communications, J. Light. Tech. **23**, 524
[22] Arfken George B and Weber Hans J 2005 *Mathematical Methods for Physicists*, Elsevier Academic Press (6th edition, Ch. 1, p. 48).
[23] Gloge D 1971 Dispersion in weakly guiding fibers, Appl. Opt. **10**, 2252
[24] http://www.cargille.com/ for detailed refractive index information.
[25] Canny J 1986 Edge detection in multispectral images, IEEE Trans. Patt. Anal. Mach. Int. **8**, 679
[26] Supplemental Material, multimedia file
[27] Supplemental Material
[28] Kaplan A E, Marzoli I, Lamb W E Jr., and Schleich W P 2000 Multimode interference: Highly regular pattern formation in quantum wave-packet evolution, Phys. Rev. A **61**, 032101 (2000)
[29] Chen C-L 2007 *Foundations for Guided-Wave Optics*, Wiley Interscience (Ch. 9, p. 233)
[30] Lamb R N, Furlong D N 1982 Controlled wettability of quartz surface, *J. Chem. Soc., Faraday Trans. 1*, **78**, 61-73